\documentclass[%
reprint,
superscriptaddress,
showpacs,
amsmath,amssymb,
aps,
prb,
final,
]{revtex4-1}
\usepackage[pdftex]{graphicx}
   \graphicspath{{.}}
\usepackage{epstopdf} 
\usepackage{subfig}
\usepackage{footmisc}
\usepackage{verbatim}
\usepackage[svgnames]{xcolor}

\begin{document}

\title{First-principles study of luminescence in Eu$^{2+}$-doped inorganic scintillators}
\author{A. Chaudhry}
\affiliation{Lawrence Berkeley National Laboratory, One Cyclotron Rd., Berkeley, CA 94720}
\affiliation{Department of Applied Science, University of California, Davis, CA 95616}
\author{R. Boutchko}
\affiliation{Lawrence Berkeley National Laboratory, One Cyclotron Rd., Berkeley, CA 94720}
\author{S. Chourou}
\affiliation{Lawrence Berkeley National Laboratory, One Cyclotron Rd., Berkeley, CA 94720}
\author{G. Zhang}
\affiliation{Lawrence Berkeley National Laboratory, One Cyclotron Rd., Berkeley, CA 94720}
\affiliation{Department of Applied Science, University of California, Davis, CA 95616}
\author{N. Gr{\o}nbech-Jensen}
\affiliation{Lawrence Berkeley National Laboratory, One Cyclotron Rd., Berkeley, CA 94720}
\affiliation{Department of Applied Science, University of California, Davis, CA 95616}
\affiliation{Department of Chemical Engineering and Materials Science, University of California, Davis, CA 95616}
\affiliation{Department of Mechanical and Aerospace Engineering, University of California, Davis, CA 95616}
\author{A. Canning}
\affiliation{Lawrence Berkeley National Laboratory, One Cyclotron Rd., Berkeley, CA 94720}
\affiliation{Department of Applied Science, University of California, Davis, CA 95616}
\affiliation{Department of Chemical Engineering and Materials Science, University of California, Davis, CA 95616}

\date{\today}

\begin{abstract}
Luminescence in Eu$^{2+}$ activated materials corresponds to a transition from an excited state where the lowest  Eu {5\textit d} level is filled with one electron (often called the (Eu$^{2+}$)$^*$ state)  to the ground state with half-filled {4\textit f} shell 
with seven electrons of the same spin. 
We have performed theoretical calculations based on Density Functional Theory to determine the ground state band structure of Eu-doped materials as well as study the (Eu$^{2+}$)$^*$ excited state. 
Calculations were performed on Eu doped materials, experimentally known to be either scintillators or non-scintillators, in order to relate theoretically calculable parameters to experimentally observed properties.
Applying criteria previously developed for Ce-doped systems (A.Canning, A. Chaudhry, R. Boutchko and N. Gr\o{}nbech-Jensen, Phys. Rev. B {\bf 83} 125115(2011)) to new Eu-doped materials we developed a list of candidate materials for new bright Eu activated scintillators. Ba$_2$CsBr$_5$:Eu is an example of a new bright scintillator from our candidate list that has been synthesized in microcrystalline powder form.
As discussed in our previous paper on Ce-doped materials this approach was designed as a systematic high-throughput method to aid in the discovery of new bright scintillator materials by prioritization and down-selection on the large number of potential new materials.
\end{abstract}


\maketitle 

\section{\label{sec:intro}Introduction}

Europium (Eu) is one of the few lanthanides that commonly exists in stable divalent and trivalent states. This behavior is unusual for most lanthanides, which almost exclusively form compounds with an oxidation state of +3. In the case of Eu the +2 state is stabilized by the fact that it corresponds to a half-filled $4f$-shell. Eu$^{2+}$ is a frequently used dopant in luminescent materials because of the dipole allowed optical transition from the lowest 4\textit f$^6$5\textit d$^1$ excited state to the 4\textit f$^7$5\textit d$^0$ ground state. In terms of its use as an activator for scintillators or X-ray phosphors the fact that Eu can exist in both the Eu$^{2+}$ and Eu$^{3+}$ form makes the Eu$^{2+}$ ion a good hole trap for the holes created by the incident gamma or X-ray. This is similar to Ce which acts as a hole trap in the Ce$^{3+}$ valence state since the Ce$^{4+}$ form is also stable.

The scintillation decay time of 400-1500ns for Eu$^{2+}$ activated compounds limits their use for applications requiring ultrafast scintillation such as medical tomography. However, the decay time is fast enough to avoid deterious effects of signal pile up for many other application areas such as homeland security, non-proliferation etc. The recent "rediscovery" of SrI$_2$:Eu as an extremely bright scintillator with energy resolution similar to that of LaBr$_3$:Ce has sparked interest in investigating Eu doped compounds to be employed for radioisotope identification.\cite{wilson:707917,cherepy2009} More recently two new scintillators, Ba$_2$CsI$_5$:Eu and BaBrI:Eu, with light yield comparable to SrI$_2$:Eu have been discovered\cite{bourret2009eu2,bourret2010babri} although BaBrI:Eu was previously known to be a bright X-ray phosphor.\cite{rajan2005x} In particular, Ba$_2$CsI$_5$:Eu is reported to be less hygroscopic than LaBr$_3$ and SrI$_2$.\cite{bourret2009eu2} Furthermore, in a recent review\cite{dorenbos2010fundamental} Dorenbos observes that significant improvements in scintillation light yield over SrI$_2$:Eu may be possible for Eu$^{2+}$ activated scintillators. 

First-principles calculations within the framework of density functional theory (DFT) have previously been  employed to study and search for new bright rare-earth doped inorganic scintillator materials.\cite{klintenberg2002potential,andriessen2007lattice,ortiz2009data,Setyawan2011} We have previously developed a systematic calculation procedure to study Ce$^{3+}$ doped scintillator materials based on studies of about a hundred inorganic host compounds.\cite{Canning2011} Our theoretical criteria are based on calculating the relative positions of Ce 4\textit f and 5\textit d states with respect to host valence and conduction band edges, respectively. 
This approach has been validated for known Ce scintillators, non-scintillators and new Ce-doped candidate materials for bright scintillation have also been predicted.\cite{boutchko2009cerium,chaudhry2009first,chaudhry2011first} 
Theoretically calculable parameters that we use to predict candidate materials for bright scintillation are 
\begin{enumerate}
\item
The size of the host material bandgap. The number of electron-hole pairs produced by the incident $\gamma$-ray is inversely proportional to the host material bandgap. 
\item
The energy difference between the valence band maximum (VBM) of the host and the Eu {4\textit f} level. The Eu {4\textit f} level must be above the VBM for scintillation. 
\item
The level of localization of the lowest {\textit d} character excited state 
determines if it is a host conduction band (CB) state or a Eu {5\textit d} character state. This is to determine if the excited Eu {5\textit d}  state is in the gap of the host material.
\end{enumerate}

Overall a necessary condition for luminescence and scintillation is that the Eu {4\textit f} and {5\textit d} levels should lie in the gap of the host material and the Eu {4\textit f}-VBM gap should not be too large or hole trapping on the Eu site will be less favored. The excited Eu {5\textit d} level should not be too close to the CBM or thermal quenching will reduce the brightness at room temperature. Trapping processes on the host associated with electron traps, hole traps or self-trapped excitons can also reduce or quench the brightness of a scintillator but they are beyond the scope of the present study.   
The present work is an extension of our calculation scheme for Ce-doped systems to study Eu luminescence for candidate scintillator materials and the reader should consult our previous paper \cite{Canning2011} for more details of the approach. 
We have previously reported preliminary theoretical results for Ba$_2$CsI$_5$:Eu and BaFI:Eu.\cite{bourret2009eu2,gundiah2010scintillation} In this paper we report on more accurate calculations for those systems, using larger cell sizes, as well as new Eu doped candidate materials. 

This theoretical approach can also be used to select candidate materials for Eu$^{2+}$-activated phosphors for efficient lighting applications etc. noting that the necessary criterion for such applications is that the Eu {4\textit f} and {5\textit d} levels should lie in the gap of the host material and the {5\textit d} should not be too close to the CBM to prevent thermal quenching. Unlike scintillators, in phosphors used for lighting, the {4\textit f} electron is directly excited, so the size of the bandgap and the proximity of the {4\textit f} to the VBM are not directly related to a phosphor's performance for these types of applications.  


\section{Calculation Details }
\label{section:calculation details}
In order to simulate the effect of a single Eu$^{2+}$ dopant in a host lattice we construct a large supercell by periodically repeating the unit cell of the host crystal and then replace 
one of the host divalent cations (Ba$^{2+}$, Sr$^{2+}$ or Ca$^{2+}$) with Eu$^{2+}$. The initial atomic positions and symmetry information of the host crystal were taken from the Inorganic Crystal
Structure Database (ICSD).\cite{ICSD} 
Atomic relaxation studies of the doped system were performed within the framework of density functional theory using the VASP code.\cite{KRESSE:1993kq,Kresse:1996kf,Kresse:1996cu}
Spin-polarized calculations were performed with PBE\cite{Perdew:1996oq} and LDA approximations to the exchange-correlation functional and using the frozen-core projector-augmented wavefunction (PAW) method\cite{blochl1994projector} as implemented in the VASP code.\cite{kresse1999ultrasoft}  The Europium pseudopotential
was chosen to include (5$s$,5$p$,6$s$,4\textit f,5\textit d) as valence electrons. The plane wave cutoff energy for the electronic wavefunctions was set to 500 eV. Integration within the Brillouin zone was performed on a $\Gamma$ centered grid of \textit{k}-points. The number of irreducible \textit{k}-points was chosen to be 4 or 8 depending on the size and geometry of the supercell. The total energy convergence criterion was set to 10$^{-6}$~eV and the maximum component of force acting on any atom in the relaxed geometry was less than 0.01~eV$/ {\mathrm \AA}$.  

PBE+U\cite{anisimov1997first} calculations were performed using the rotationally invariant method of Dudarev\cite{dudarev1998electron} 
for an on-site +U correction to treat the Eu {4\textit f} electrons with a single parameter U$_{\rm eff}(= U - J)$. We tuned the 
empirical U$_{\rm eff}$ parameter to give the best agreement with experimental data and related previous calculations for the ground state Eu {4\textit f} to host VBM gap as described in the next section. 

Host bandgaps were calculated at the level of PBE, HSE functionals and the GW\cite{hedin1965new} approximation. Commonly used density functionals such as PBE and LDA are known to underestimate the bandgap of semiconductors and insulators. A hybrid functional approach combining a fraction of screened exchange with an explicit density functional has been shown to give eigenenergies which are generally in much better agreement with experiment; especially for semiconductors.\cite{heyd2003hybrid,heyd2006erratum} These hybrid functional eigenvalues and eigenfunctions are therefore superior starting points for quasi-particle corrections using the GW approximation. 
Bandgaps calculated using an HSE+G$_0$W$_0$ approach have been shown to be in  good agreement with experiments.\cite{fuchs2007quasiparticle} HSE06\cite{heyd2003hybrid,heyd2006erratum} calculations were performed using the default fraction ($\alpha$=0.25) of nonlocal Fock exchange. Subsequently, quasi-particle bandgaps were determined within the single-shot G$_0$W$_0$ approach. Convergence of representative G$_0$W$_0$ calculations were checked with respect to the number of empty bands and energy cutoffs used in the GW calculation. 

\section{\label{sec:results}Results and Discussion}

\subsection{Determination of U$_{\rm eff}$ parameter}

Eu$^{2+}$ in the ground state, has a half-filled 4\textit f$^7$ shell with all the spins aligned and in the excited state has the 4\textit f$^6$5\textit d$^1$ structure with the six 4\textit f electrons spin aligned leaving one empty 4\textit f state. While for a very accurate modeling of the 4\textit f electrons and their interactions more advanced theories than PBE+U may be required we found that a +U correction for Eu doped materials gives reasonable quantitative agreement between theory and experiment for 4\textit f electron energy levels. Also our purpose in this paper is to provide a high throughput method for qualitative prediction of scintillator properties of new materials rather than perform costly more accurate calculations for a small number of systems. In order to set the correct ground state spin alignment of the 4\textit f electrons we set NUPDOWN=7 in the VASP input file. 

Typical values of U$_{\rm eff}$ reported in the literature for bulk Eu$^{2+}$ compounds are $\geq$ 6eV.\cite{larson2006electronic,shi2008electronic} We found that setting U$_{\rm eff}$= 6eV places the Eu$^{2+}$ 4\textit f level incorrectly with respect to the host valence band maximum. For example, PBE+U calculations of BaI$_2$:Eu position the Eu$^{2+}$ 4\textit f level below the VBM (Figure 1). This is inconsistent with the fact that BaI$_2$:Eu is a known scintillator material and hence the Eu 4\textit f states should lie above the host VBM.\cite{cherepy2008strontium} 
From our studies of Ce$^{3+}$ doped compounds we found that the 4\textit f states of rare-earth dopants can have very atomic-like character as opposed to an itinerant nature which is possible in Ce bulk compounds. Therefore, the +U correction required for wide bandgap Eu doped systems may be significantly different from bulk Eu systems and  
we explored the possibility of determining U$_{\rm eff}$ empirically based on our previous studies of Ce-doped systems as well as experimentally measured 4\textit f-VBM energy gaps for Eu$^{2+}$ doped inorganic compounds.


\begin{figure}
\centering
\includegraphics[scale=0.5,trim=50 0 0 20, clip=true]{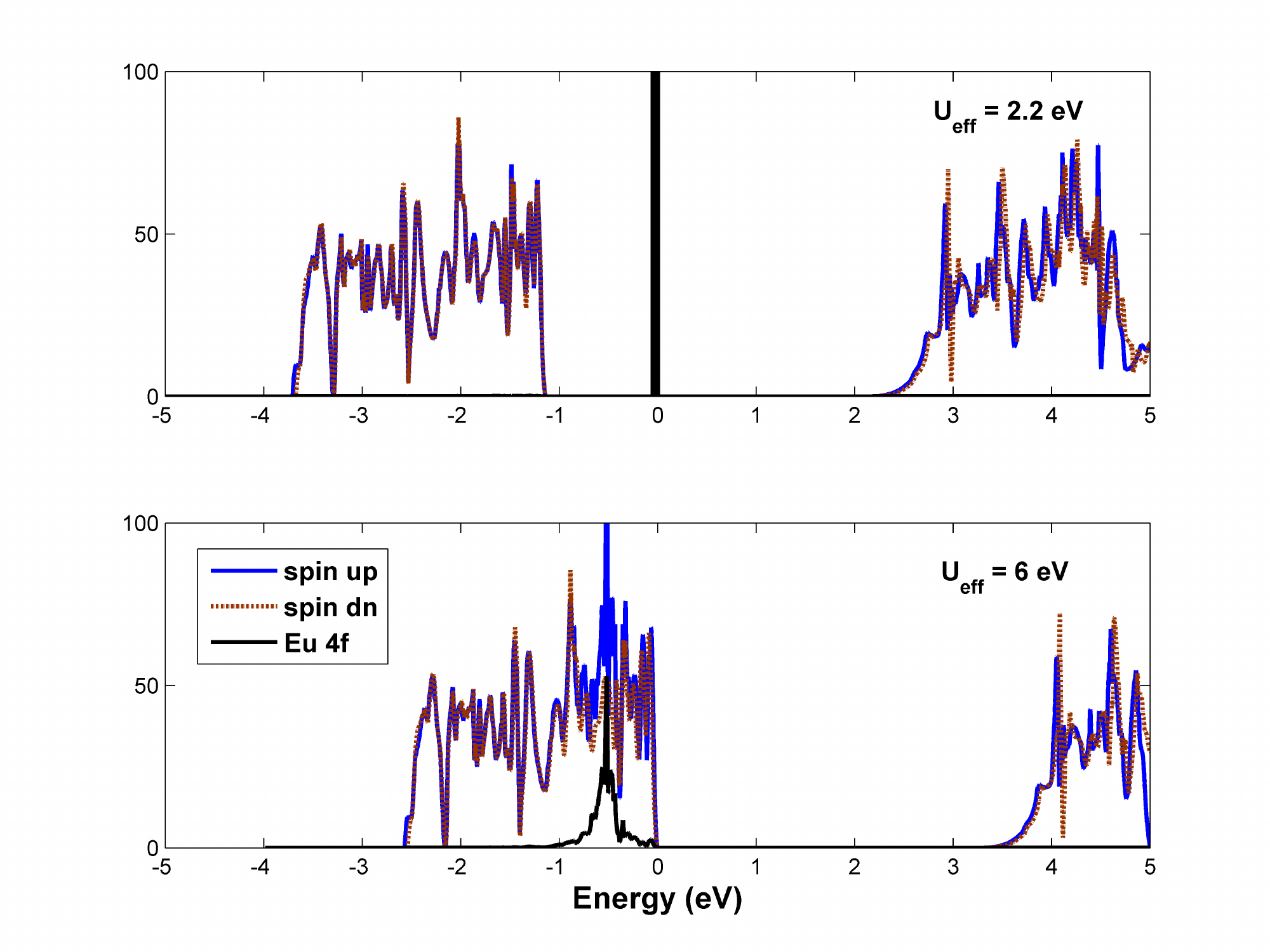}
\captionsetup{justification=justified}
\caption{\label{fig:fig1} 
Density of states calculated for BaI$_2$:Eu with (a) U$_{\rm eff}$= 2.2eV and (b) U$_{\rm eff}$= 6eV. Eu 4$f$ (spin up) states are clearly below the VBM for U$_{\rm eff}$= 6eV. Fermi level is set at 0eV.}
\end{figure}

Direct experimental measurement of the Eu$^{2+}$ 4\textit f-VBM gap is available for a very few compounds such as CaGa$_2$S$_4$:Eu.\cite{hiraguri2006photoconductivity} The energy level scheme of large bandgap compounds such as CaF$_2$, SrF$_2$ and BaF$_2$ doped with Eu$^{2+}$ is known from detailed experiments. CaF$_2$, SrF$_2$ and BaF$_2$ have bandgaps of 12.1eV, 11.25eV and 11.0eV respectively,\cite{rubloff1972far} with estimates for the 4\textit f-VBM energy gap at 7.9eV, 7.5eV and 7.0eV respectively.\cite{pedrini2007photoionization} 
The emphasis in our search for new bright scintillator materials is not on such materials since a large band gap fundamentally limits the number of electron-hole (e-h) pairs generated by an incoming $\gamma$-ray. Also in the context of the bandgap error in LDA calculations for these systems, since the 4\textit f level can lie above the midpoint of the bandgap, we may expect a large error in our 4\textit f-VBM due to the bandgap error. Therefore, fitting the +U parameter to these systems may yield different values from the lower bangap systems we are interested in where the Eu$^{2+}$ 4\textit f level is close to the VBM. For the lower bandgap systems we may expect the error in the 4\textit f level due to the bandgap error to be small and the self-interaction error to be dominant. We, therefore, only used the experimentally determined 4\textit f-VBM gap (=1.75 eV\cite{hiraguri2006photoconductivity}) for CaGa$_2$S$_4$:Eu to fit the U$_{\rm eff}$ parameter where the 4\textit f level is relatively close to the VBM.  

Dorenbos has proposed an alternative empirical method to estimate the location of Eu$^{2+}$ levels relative to the host valence and conduction band edge.\cite{dorenbos2003energy} The 4\textit f-VBM gap is estimated based on the energy required for charge-transfer from the host VB to Eu$^{3+}$ while the location of the lowest 5\textit d level requires an estimation of the host conduction band edge. This approach, as stated in the paper,\cite{dorenbos2003energy} can introduce systematic errors as large as 0.5eV and it depends on the measured Eu absorption and emission wavelengths.\cite{Dorenbos2003}  Therefore we did not use this data for fitting the U$_{\rm eff}$ parameter.

We estimate the 4\textit f-VBM gap using the same U$_{\rm eff}$ for Eu 4\textit f electrons as determined for Ce 4\textit f electrons in our earlier work.\cite{Canning2011} We verified that the character of the Eu 4\textit f electrons is very similar to the Ce 4\textit f electron and atomic-like in nature. Table~\ref{tab:table1} summarizes these results for selected Eu-activated luminescent materials. 
Choosing U$_{\rm eff}$=2.2eV gave very good agreement with experimental data for CaGa$_2$S$_4$:Eu. This is similar to the U$_{\rm eff}$ used for lower band gap and more covalent iodides and sulfides in our studies of Ce-doped systems.\cite{Canning2011} We set U$_{\rm eff}$ = 2.5eV for oxides and non-iodide halides as was found to give the best fit to experimental data for Ce systems.  
Since the 4\textit f electrons are screened by the outer 5\textit s and 5\textit p electrons the character of the Eu 4\textit f electrons remains very similar in different host compounds so we can use the  U$_{\rm eff}$ parameter thus determined to estimate the Eu 4$f$-VBM energy gap in new candidate host compounds from first-principles. While the limited amount of experimental data for the 4\textit f level in Eu doped systems does not allow us to directly confirm this from the experimental point of view we did find this to be the case for Ce doped systems where experimental data for a range of different host materials with different 4\textit f levels is available.\cite{Canning2011}

\begin{table*}
\caption{\label{tab:table1}Calculated band gaps and relative {4\textit f}
and {5\textit d} levels for Eu doped compounds. Energies are given in eV. All the listed host compounds have a direct gap 
except Sr$_2$MgSi$_2$O$_7$ where the indirect gap is listed. The experimental bandgaps quoted are the optical bandgap which for this material from the PBE bulk bandstructure is about 0.3eV above the indirect gap. The {4\textit f}\textendash VBM level for CaGa$_2$S$_4$ is the experimental value used to fit the U$_{\rm eff}$ value for the theoretical calculations of the other {4\textit f}\textendash VBM levels in the table.}
\begin{ruledtabular}
\begin{tabular}{lcclclcl}
 {\bf Compound}&
 \multicolumn{3}{c}{\bf Band gap}&{\bf Eu {4\textit f}\textendash VBM}&{\bf {4\textit f} $\rightarrow$ {5\textit d}}&{\bf Eu {5\textit d}\textendash CBM}\\
 {}&{PBE}&{HSE06}&{HSE06+G$_0$W$_0$}\footnote{Values in parentheses refer to experimental bandgaps quoted from the literature.}&{}&{}&(before Stokes shift)\\
 \hline
SrS & 2.7  & 3.44& 4.63 (4.32\cite{Chartier2006}) & 1.75 (U$_{\rm eff}$=2.2) & 2.25\cite{Smet2006} & $\sim$0.6
\\
Sr$_2$MgSi$_2$O$_7$ & 4.7 &6.49 & 7.36 (7.1\cite{holsa2009electronic},7.45\cite{Chen2006}) & 3.65 (U$_{\rm eff}$=2.5) & 2.78\cite{Zhang2007,Li2009}  & $\sim$0.9
\\
BaCl$_2$ & 5.06 &6.23 &7.98 & 2.8 (U$_{\rm eff}$=2.5)  & 3.29\cite{Brixner1976,Kobayasi1980} & $\sim$1.9
\\
BaBr$_2$ & 4.27 &5.45 &6.78 & 1.9 (U$_{\rm eff}$=2.5) & 3.35\cite{Iwase1994,Troster2002} & $\sim$1.4
\\
BaI$_2$  & 3.33 &4.16 &5.03 & 1.1 (U$_{\rm eff}$=2.2) & 3.04\cite{cherepy2008strontium,gahane2009luminescence} & $\sim$0.8
\\
SrI$_2$ & 4.0 &4.98 &5.36 (5.5\cite{Pankratov2013},5.7-5.8\cite{Pustovarov2012})& 1.4 (U$_{\rm eff}$=2.2) & 2.95\cite{cherepy2008strontium,gahane2009luminescence} & $\sim$1.0
\\
Ba$_2$CsI$_5$ & 3.67 & 4.6 & 5.67 (5.1\cite{Janecek2011},5.3-5.5\cite{alekhin2014optical})  & 1.2,1.4 (U$_{\rm eff}$=2.2)\footnote{Two different substitutional sites.}  & 2.94\cite{bourret2009eu2,Bizarri2011} & $\sim$0.9
\\
BaBrI & 3.43 & 4.40 & 5.39 & 1.5,1.3 (U$_{\rm eff}$=2.2, 2.5)\footnote{{4\textit f}-VBM values are for the two different choices of U$_{\rm eff}$ for this mixed halide system.}  & 3.12\cite{bourret2010babri,Bizarri2011} & $\sim$0.9
\\
BaFI & 3.98 & 4.96 & 6.27 (6.8\cite{Ohnishi2008})& 2.0 (U$_{\rm eff}$=2.5) &  3.22\cite{gundiah2010scintillation} & $\sim$1.0
\\
CaGa$_2$S$_4$  & 2.8 & &(4.4\cite{Chartier2006}) & 1.75 (U$_{\rm eff}$=2.2) & 2.36\cite{Dzhabbarov2002,Nazarov2008} & &
\end{tabular}
\end{ruledtabular}
\end{table*}

The experimental 4\textit f$\rightarrow$5\textit d absorption data in column 6 of Table~\ref{tab:table1} was provided by P. Dorenbos\cite{Dorenbos_priv} and estimated from the experimental absorption curves presented in the references listed in the table. The absorption data for BaCl$_2$, BaBr$_2$, SrS and CaG$_2$S$_4$ was previously published\cite{Dorenbos2003} while for the other systems the data was only published in graphical form and the explicit values listed by compound were not presented in the paper.\cite{Dorenbos2013}  The absorption curves for these systems often show a staircase structure as there are seven possible 4\textit f$^6$5\textit d configurations depending on which of the 4\textit f levels is empty. The results presented here are the absorption to the lowest 4\textit f$^6$5\textit d energy level.  

Except for CaGa$_2$S$_4$:Eu we did not have information on measured 4\textit f-VB gaps for compounds listed in Table~\ref{tab:table1}. Therefore, to ascertain the accuracy of calculated 4\textit f-VBM gap we used the following approach based on theory and experimental results. This is also a secondary check on our choice of U$_{\rm eff}$ as all known scintillators should have the 4\textit f and 5\textit d levels in the gap of the host material. 
First, we calculate the host bandgap for a few Eu doped luminescent compounds within the GW approximation. The calculated quasi-particle gaps are in good agreement with the experimental literature as shown in Table~\ref{tab:table1}. The GW bandgap for SrI$_2$ is in close agreement with other theoretical calculations.\cite{Erhart2013} In the second step, we estimate the 4\textit f-VBM gap from PBE+U calculations. Thereafter, subtracting the experimentally determined absorption energy of Eu {4\textit f}$\rightarrow$5\textit d(lowest) and the calculated 4\textit f-VBM gap from the quasi-particle bandgap we estimate the Eu 5\textit d-CBM gap (before the Stokes shift) for these compounds. Since Eu$^{2+}$  luminescence is observed in our chosen set of materials {4\textit f}-VBM and {5\textit d}-CBM gaps must be positive and this is indeed the case as presented in Table~\ref{tab:table1}. 

We note here that the Stokes shift of Eu$^{2+}$ in solids with (Ba, Sr and Ca) sites is on average 0.26{$\pm$}0.14eV.\cite{dorenbos2005thermal} The shift in the {5\textit d} level due to the Stokes shift must be lower than this as the Stokes shift also includes the shift in the 4\textit f compared to the ground state atomic positions. Based on significant data, Dorenbos notes that, in general, the relaxed 5\textit d state (after Stokes' shift) is located within 1.0 eV below the CBM.\cite{dorenbos2005thermal} This observation is in general agreement with our calculated unrelaxed {5\textit d}-CBM gaps in Table~\ref{tab:table1} except for the larger bandgap BaCl$_2$ system.

To illustrate the influence of local environment on the position of Eu 4$f$ we consider two systems: Ba$_2$CsI$_5$ and BaBrI. The Eu ion can occupy either 7 or 8 coordinated Ba sites in Ba$_2$CsI$_5$.\cite{bourret2009eu2} The calculated 4\textit f-VBM gap differs by a few tenths of an eV for the two cases (see Table~\ref{tab:table1}). This difference also illustrates the effect that the inner 4$f$ levels are not completely screened by the outer electrons from the effect of the crystal field of the lattice so the difference in the crystal field at the two inequivalent sites leads to different 4$f$ levels. In the case of BaBrI, substitutional Eu is 9 coordinated\cite{Gundiah2011} (4 Br$^{-}$ and 5 I$^-$ anions) with average Eu-Br bond length $\sim$0.4 {\AA} shorter than the Eu-I bond length. Thus, in this mixed halide system we used U$_{\rm eff}$=2.5eV rather than 2.2eV to determine the Eu 4$f$ level position in the host band gap since the nearest atoms to Eu are the Br. Table ~\ref{tab:table1} shows the difference in {4\textit f}-VBM for the two different choices of U$_{\rm eff}$ for BaBrI.

Two previous theoretical studies on Eu$^{2+}$ doped phosphors are relevant to our work. In their theoretical studies on Sr$_2$MgSi$_2$O$_7$:Eu, Holsa {\it et al.}\cite{holsa2009electronic} found a linear dependence of the 4\textit f-VBM gap on the choice of the U$_{\rm eff}$ parameter. However, they do not provide a systematic way to tune U$_{\rm eff}$. They also note that inclusion of spin-orbit coupling broadened the width of the occupied  Eu$^{2+}$ 4\textit f levels in the ground state from 0.2eV to 0.6eV while having a less pronounced effect on unoccupied 4\textit f states.\cite{holsa2009electronic} The experimentally observed splitting of the Eu$^{2+}$ 4\textit f ground state ($^8$S$_{7/2}$) is, however, about 0.16eV.\cite{Thole1985} Considering the associated computational complexity and incorrect broadening of occupied 4\textit f states we have not included spin-orbit coupling in our calculations. They further note that the Eu$^{2+}$ 4\textit f-VBM gap is almost identical when calculated with GGA+U or LDA+U methods.\cite{holsa2009electronic} We also found that positioning of the occupied Eu$^{2+}$ 4\textit f states is largely insensitive to the choice of functional. 

Another study by Brito {\it et al.}\cite{brito2012dft} on BaAl$_2$O$_4$:Eu observes that using a U parameter between 4.65-7.68 eV overestimates the position of the 4\textit f level (J=0.68eV). They, however note that choosing a U parameter close to 3.0 eV ($U-J$ = 2.32eV) would give a better agreement with the experimentally estimated 4\textit f-VBM gap. This is remarkably close to the  U$_{\rm eff}$=2.5eV as determined in our systematic studies although their calculations use the FLAPW method as implemented in the WIEN2k package\cite{blaha2001} rather than the VASP code.

We have presented a systematic computational procedure to determine the position of Eu 4$f$ levels with respect to the host VBM in new compounds  which do not have large bandgaps. Therefore, our calculations can complement empirical models, such as those presented by Dorenbos.\cite{dorenbos2003energy} Our calculated Eu 4$f$ levels can be used in his model for parameter fitting, where experimental values are not available or for predictions of new materials.

\subsection{Excited state : 5d level location and localization}

A necessary condition for scintillation and luminescence is that the Eu 4$f$ and 5$d$ levels must lie in the gap of the host material. We did not find any system where Eu 4\textit f states were located below the VBM. However, precise determination of 5$d$ level location relative to the CBM is difficult due to (a) for new compounds we do not have information about Eu$^{2+}$ absorption energy and hence it is not possible to estimate the 5$d$-CBM gap as summarized for known scintillators in Table~\ref{tab:table1} and, (b) an accurate determination of the Eu 5\textit d-CBM energy gap for the (Eu$^{2+}$)* excited state is difficult using standard ground state LDA and GGA approximations to DFT. 

In our earlier work for Ce-doped luminescent materials we obtained qualitative estimates of the location of the Ce 5\textit d levels relative to the bottom of the conduction band (CBM) by measuring the localization of the first excited \textit d character state using a constrained occupancy (excited state) approach.\cite{Canning2011} A delocalized state corresponds to a conduction band state so the lowest Ce 5\textit d character state is located above the bottom of the conduction band while a localized state corresponds to the (Ce$^{3+}$)* state, therefore the occupied Ce 5\textit d state is below the CBM and scintillation is possible. In a simple model, the (Eu$^{2+}$)* excited state is composed of 6 4\textit f electrons along with an electron promoted to the lowest 5\textit d orbital. This situation is different from the (Ce$^{3+}$)* state which has no 4\textit f electron. A more accurate description of the (Eu$^{2+}$)* state may require many-body methods to accurately model the \textit{f-f} electron interactions which are not suitable for high-throughput calculations involving supercells with 80-100 atoms.

In the case of compounds where we do not have information about Eu$^{2+}$ absorption energy we calculate the relative localization of the lowest \textit d character excited state to determine if the Eu {5\textit d} is in the gap. To perform the excited state calculation we set the occupancy of the highest {4\textit f} level to zero and then the next highest energy level state will be filled. In the ground state Eu$^{2+}$ has a half filled {4\textit f} shell with all the spins aligned. The empty {4\textit f} spin down levels are typically much higher in energy than the Eu {5\textit d} or CBM so are never filled in the excited state calculation. In general, in the ground state calculation, Eu {5\textit d} states hybridize strongly with host CB states even for known scintillators so it is necessary to do the excited state calculation where the removal of a Eu {4\textit f} electron causes the Eu {5\textit d} state to drop in energy and move into the gap. 
\color{Black}Even for these excited state calculations in many cases there is still some hybridization of the Eu {5\textit d} state with CB states as well as finite size effects and for these reasons as well as the simple nature of our constrained occupancy excited state calculation we do not expect accurate 5\textit d-CBM values from these calculations. These issues are discussed in more detail for the case of Ce doped systems in our previous paper.\cite{Canning2011}   
\color{Black}
As in the case of the Ce studies we do not use a +U correction for the Eu {5\textit d} states as we expect the self-interaction error to be small for {5\textit d} states as they are much less localized than {3\textit d} states for transition metals where a +U correction is typically used. For all the scintillators we report in this work the CBM is of mainly {\textit d} character as well, {5\textit d} in the case of Ba, so we expect there to be some cancellation of the self-interaction energy error for the 5\textit d-CBM value. 

\begin{table*}
\caption{\label{tab:table2}Calculated PBE bandgaps, {4\textit f}\textendash 
VBM energy gap and localization of the (Eu$^{2+}$)$^{*}$ excited state for Eu-doped compounds. 
Experimental luminosity data is taken from Ref.~[\onlinecite{[Comprehensive database of 
scintillation properties of inorganic materials. ] scintillatorLBL}] 
and  references therein except for Sr$_2$CsI$_5$ which is from Ref.~[\onlinecite{patent2012}]. Asterisks (**) indicates no observed Eu$^{2+}$ emission. 
Predictions based on our calculations are marked with $\dagger$.
} 
\begin{ruledtabular}
\begin{tabular}{lccccl}
 {\bf Compound}&
 {\bf PBE Bandgap}&{\bf Eu {4\textit f}\textendash VBM}&\multicolumn{2}{l}{\bf (Eu$^{2+}$)$^{*}$ localization}&{\bf Luminosity}\\
 {(atoms in supercell)}&{(eV)}&{(eV)}&{\%}&{ratio}&{(photons/MeV)}\\
 \hline
Ba$_2$Ga$_2$S$_5$ (72) & 2.7 & 1.3 & 1.07 & 0.06 & ** 
\\
SrZrO$_3$ (40) & 3.5 & 2.6 & 0.16 & 0.02 & **
\\
SrGa$_2$O$_4$ (56) & 3.35 & 2.97 & 1.6 & 0.66 & **
\\
BaS (64) & 2.2 & 1.63 & 4.2 & 1.2 & **
\\
$\dagger$BaO (32) & 2.3 & 1.8 & 1.2 & 0.08 & 
\\
$\dagger$Ba$_2$HfS$_4$ (56)& 0.79 & 0.52 & 1.5 &0.09 & 
\\
BaCl$_2$ (48)& 5.06 & 2.8   & 43.8 & 6.83 & 52,000
\\
BaBr$_2$ (48) & 4.27 & 1.9 & 29.4 & 4.6 & 58,000
\\
BaI$_2$ (48) & 3.33 & 1.1 & 22.2 & 4.0 & 40,000
\\
BaFI (48)& 3.98 &  2.0 & 51.1 & 5.1 & 55,000 
\\
BaBrCl (192) & 4.45 & 2.45  & 42.5 & 25 & 52,000
\\
BaBrI (48)& 3.43 &1.3  & 33.2 & 5.26 & 81,000
\\
SrI$_2$ (96)& 4.0 & 1.4 & 67.3 & 14.3 & 120,000
\\
CaI$_2$ (54)& 3.46 & 1.4  & 38.5 & 5.0 & 110,000
\\
Sr$_2$CsI$_5$ (32)& 3.95 & 1.3   & 28.0 & 4.77 & 56,000
\\
Ba$_2$SiO$_4$ (56)& 5.0 & 2.3 & 45.3 & 7.25 & 40,000
\\
BaSiO$_3$ (40) & 5.6 & 3.1 & 16.1 & 4.14 & 8,000
\\
Ba$_3$P$_4$O$_{13}$ (80) & 5.6 & 3.4   & 24.4 & 16.67 & 25,000
\\
Sr$_2$MgSi$_2$O$_7$ (48) & 4.7 & 3.65   & 18.9 & 6.1 & 4,000
\\
Ba$_2$MgSi$_2$O$_7$ (48) & 4.4 & 3.1   & 24.5 & 3.7 & 10,000
\\
Ba$_5$SiO$_4$Br$_6$ (64) & 4.4  & 2.4  & 18.5 & 7.5 & \footnote{Ba$_5$SiO$_4$Br$_6$, BaAl$_2$S$_4$ and SrS are known Eu activated phosphors although there is no published data on their scintillation properties. \label{first_footnote}}
\\
BaAl$_2$S$_4$ (84)& 3.73 & 1.56 &  50.0 & 16.6 & \footref{first_footnote}
\\
SrS (64)& 2.67 & 1.6 &  14.9 & 3.63 & \footref{first_footnote}
\\
$\dagger$Ba$_2$CsI$_5$ (64)& 3.67 & 1.4   & 43.6 & 6.57 & 97,000
\\
$\dagger$Ba$_2$CsBr$_5$ (64)& 4.6 & 2.2   & 34.7 & 5.27 & 91,800
\\
$\dagger$Sr$_4$OI$_6$ (44) & 4.1  & 1.3   & 23.1 & 5.0 & 
\\
$\dagger$BaCsB$_3$O$_8$ (66)& 5.8 & 3.4 &18.7 & 2.7 & {}
\end{tabular}
\end{ruledtabular}
\end{table*}

Table~\ref{tab:table2} presents the results of host PBE bandgaps, ground state Eu 4\textit f-VBM gaps for the relaxed structure and localization percentage and ratio of the (Eu$^{2+}$)* excited state for a selection of scintillators and non-scintillators. The localization percentage in Table~\ref{tab:table2} is the percentage of the normalized single electron density in a Voronoi cell centered on the Eu atom.  The ratio in Table~\ref{tab:table2} is the ratio of the localization percentage of a state on the Eu site to its next largest localization percentage on a different cation. Non-scintillators refers to compounds where Eu$^{2+}$ luminescence has not been observed experimentally.
We have performed calculations for about sixty different Eu doped materials and present the best candidates for new bright scintillation, based on our calculated criteria, at the bottom of Table~\ref{tab:table2} as well as Ba$_2$HfS$_4$:Eu and BaO:Eu which are predicted non-scintillators. We obtain good qualitative agreement for known scintillators and non-scintillators for the relation between our calculated parameters and their performance as scintillators. In particular, there is no localization of the excited states for non-scintillators while known bright scintillators such as SrI$_2$:Eu and BaBrI:Eu have moderate bandgap, low 4$f$-VBM gap, and very localized 5\textit d character states centered on Eu. 
Convergence tests were carried out with respect to cell size for the systems in Table~\ref{tab:table2}. The cell sizes quoted in Table~\ref{tab:table2} were chosen to give the 4$f$ levels converged to less than 0.1eV  and the localization data converged to within 10\%.

Amongst the non-scintillators, the lowest excited state has predominately $d$ character in BaS, BaO, SrZrO$_3$ and Ba$_2$HfS$_4$ corresponding primarily to Ba 5$d$, Zr 4$d$ and Hf 5$d$ character states. Whereas in SrGa$_2$O$_4$ and Ba$_2$Ga$_2$S$_5$ the excited state has strongly Ga 4$s$ character. Thus the presence of the second cation (Zr, Hf or Ga) introduces a lower energy state which constitutes the CBM. The CBM state associated with the second cation is below the Sr or Ba $d$ character states as well as the first Eu $5d$ state.
In all our studies for Eu doped materials we found that if the CBM has no $d$ character then there is no Eu emission. If the CBM is of $s$ or $p$ character then it seems to put in a relatively much lower CBM so that the Eu 5$d$ states are always in the conduction band. Therefore determining if the dominant character of the CBM of the host material is $d$ or not can possibly be used to determine if the doped material is a candidate phosphor or scintillator  with Eu doping. Furthermore, we also find that even if the CBM is of $d$ character unless that $d$ character is associated with the dopant site atom for Eu (Ca,Sr and Ba in our studies) then the Eu 5$d$ is in the conduction band and there is no emission. We also find that for many systems there is some level of hybridization between the Ca,Sr and Ba 5d states and the other cation states forming the CBM. Further studies would be required to see if these rules hold for all Eu and Ce doped phosphors.  
Dorenbos \cite{dorenbos2005thermal} distinguishes between two types of Eu phosphor materials which he refers to as type I and type II. For type I the CBM is formed by $d$ character states from the Ca, Sr or Ba while for type II the CBM has the character of one of the other cations in the host material. As mentioned above, in our studies we find there are many materials that fall between these two classifications in the sense that the CBM is formed by a hybridization of states from the different cations of the host material. Overall though we find the bright scintillators have a CBM of predominantly Ca, Sr and Ba $d$ character (type I) while the weaker scintillators and non-scintillators are closer to type II systems, but there exist also non-scintillators of type I such as BaO.    

BaS crystallizes in the rocksalt $Fm\overline3m$ structure similar to CaS and SrS both of which are known to exhibit Eu$^{2+}$ activated luminescence. There are contradictory reports in the literature on Eu$^{2+}$ emission in BaS:Eu with earlier data of Kasano et al.\cite{Kasano1984} quoting an emission wavelength at 572nm. Recently, Smet et al.\cite{Smet2006} observed a peak centered at 873nm characterized as "anomalous emission" but no band at 572nm in the emission spectra. They present an energy level scheme for BaS:Eu$^{2+}$ but state that placement of 5\textit d states is qualitative. Let us try to estimate the 5$d$-CBM gap noting that the reported low temperature {4\textit f}$\rightarrow${5\textit d}(lowest) absorption wavelength is 542nm.\cite{Smet2006} The theoretically calculated 4$f$-VBM gap in BaS:Eu is 1.63 eV. The bandgap of BaS from the optical absorption spectrum is reported as 3.88eV\cite{Zollweg1958} and 3.9eV.\cite{Saum1959} Subtracting the absorption energy and 4$f$-VBM energy from the bandgap gives the lowest 5\textit d state slightly above the CBM. On the other hand, if we consider the bandgap of 3.49eV as quoted by Smet et al.\cite{Smet2006} then the location of the 5$d$ level is $\sim$0.5eV above the CBM, in very good agreement with their results. Overall our results based on measuring localization of the excited (Eu$^{2+}$)* state are in agreement with recent experimental results for BaS:Eu.\cite{Smet2006} 

Similar to BaS, BaO crystallizes in the rocksalt structure and based on negligible localization of the excited state on Eu site we predict it to be a non-scintillator.  We are not aware of any published data on absorption (or emission) wavelength for BaO:Eu and there is one publication where it was studied for potential scintillation applications however, the host was found to be excessively hygroscopic.\cite{cherepy2009scintillators}
It is instructive to note that MO and MS (M= Ba, Sr, Ca) compounds, by Dorenbos' classification, are examples of type I systems where the band gap decreases much faster than Eu emission (and absorption) energies.\cite{dorenbos2005thermal} 

Alkaline earth halide and mixed halide compounds are some of the brightest known Eu-activated scintillators and our calculations qualitatively agree with measured data in the literature for this class of systems. The barium systems BaCl$_2$, BaBr$_2$, BaI$_2$, BaClBr and BaBrI are isomorphous and occur in the orthorhombic PbCl$_2$ structure. The bandgap and 4$f$-VBM gap increases as we go up the periodic table from diiodides to dichlorides. Barium fluorohalides such as BaFI occur in a tetragonal structure similar to matlockite (PbFCl). BaFI is not hygroscopic unlike other Ba halides or mixed halides however, due to its layered structure it cleaves easily.\cite{gundiah2010scintillation} BaFI:Eu, BaBrCl:Eu and BaBrI:Eu are known X-ray phosphors with BaFI:Eu being the most commonly used X-ray storage phosphor. As can be seen from Table II all the alkaline halides have very localized excited Eu 5d states below the CBM while the brightest ones (BaBrI, SrI$_2$) also have the lowest bandgaps and 4$f$-VBM gaps. BaI$_2$  has the lowest bandgap of all the alkaline earth halides in the table but is less bright than BaBrI and SrI$_2$. This may be due to trapping centers as suggested in some of the experimental studies of BaI$_2$.\cite{cherepy2008strontium}  The alkaline earth diiodide CaI$_2$ is one of the brightest known scintillators and occurs in the hexagonal $P\overline3m1$, structure which is different from the other diiodides like SrI$_2$ and BaI$_2$ which are orthorhombic. This in part explains why the calculated bandgap for CaI$_2$ is actually lower than for SrI$_2$. From the theory we find it has a low 4$f$-VBM and a very localized excited Eu 5$d$ state.  
The oxide based hosts listed in Table~\ref{tab:table2} are characterized by higher bandgaps than the halides or sulphides, larger 4$f$-VB gaps and relatively good localization on the Eu site although typically not as high as the bright halides. Many of the oxide host materials such as the silicates are relatively easy to grow in crystal form and have higher thermal stability so even though no oxide based scintillator is known that has brightness comparable to the very bright halide scintillators such as SrI$_2$ there is continued interest in new oxides for phosphor as well as scintillator applications. Ba$_2$SiO$_4$:Eu is a well known bright phosphor used for solid-state lighting and as a scintillator its luminosity has been recently measured at 2.7 times that of BGO.\cite{Eagleman2012} Orthorhombic $Pnam$ Ba$_2$SiO$_4$ has a lower bandgap, lower 4$f$-VB as well as higher localization as compared to the less known BaSiO$_3$ and hence qualitatively we would expect Ba$_2$SiO$_4$ to be brighter which agrees with experiments.\cite{Eagleman2012} Persistent luminescent materials Sr$_2$MgSi$_2$O$_7$:Eu and Ba$_2$MgSi$_2$O$_7$:Eu have recently received attention as storage phosphors. Calculations were done for the tetragonal $P\overline42{_1}m$ phase of Sr$_2$MgSi$_2$O$_7$\cite{holsa2009electronic} and the monoclinic $C2/c$ phase of Ba$_2$MgSi$_2$O$_7$ which were consistent with scintillation measurements.\cite{Derenzo2011} The estimated 4$f$-VB gap is lower for Ba$_2$MgSi$_2$O$_7$:Eu whereas the localization on the Eu site is similar. 

Sulphides are attractive hosts for phosphor applications since they have small bandgaps so have the potential to produce very bright scintillators although their experimentally measured luminosities have been well below those of the halides. BaAl$_2$S$_4$:Eu is a well known bright blue phosphor.\cite{Miura1999} The calculated 4$f$-VB gap as well as the excellent localization of the excited $d$ state on the Eu site are indicative of a good candidate host for bright Eu$^{2+}$ emission in agreement with experimental data. Similar agreement is also found for SrS:Eu. These materials are thus potential candidates for Eu-activated scintillation.


\begin{figure*}
\subfloat[]{\label{fig:2a}\includegraphics[scale=0.37]{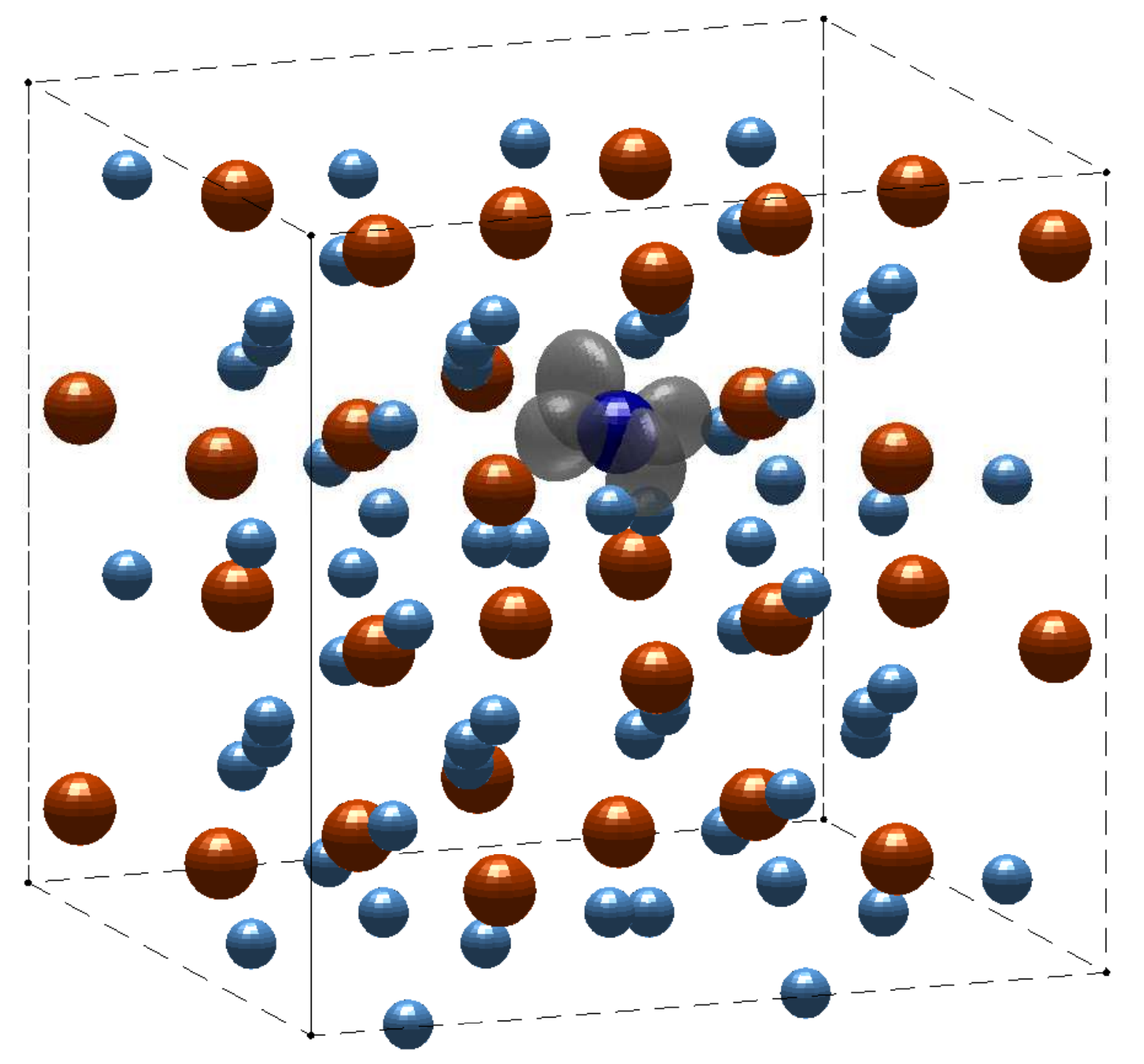}} \qquad
\subfloat[]{\label{fig:1c}\includegraphics[scale=0.38]{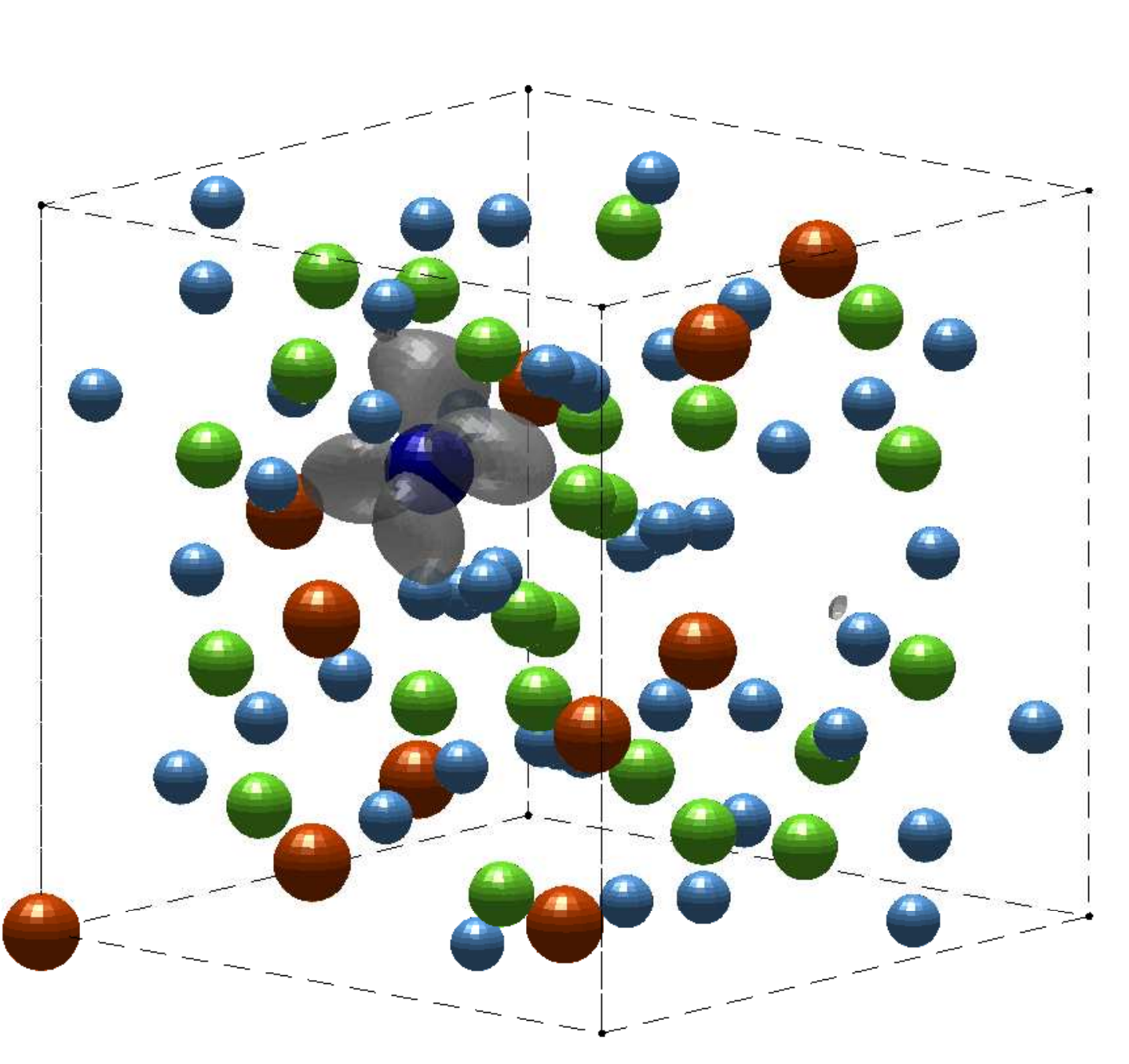}} \qquad
\subfloat[]{\label{fig:2c}\includegraphics[scale=0.39]{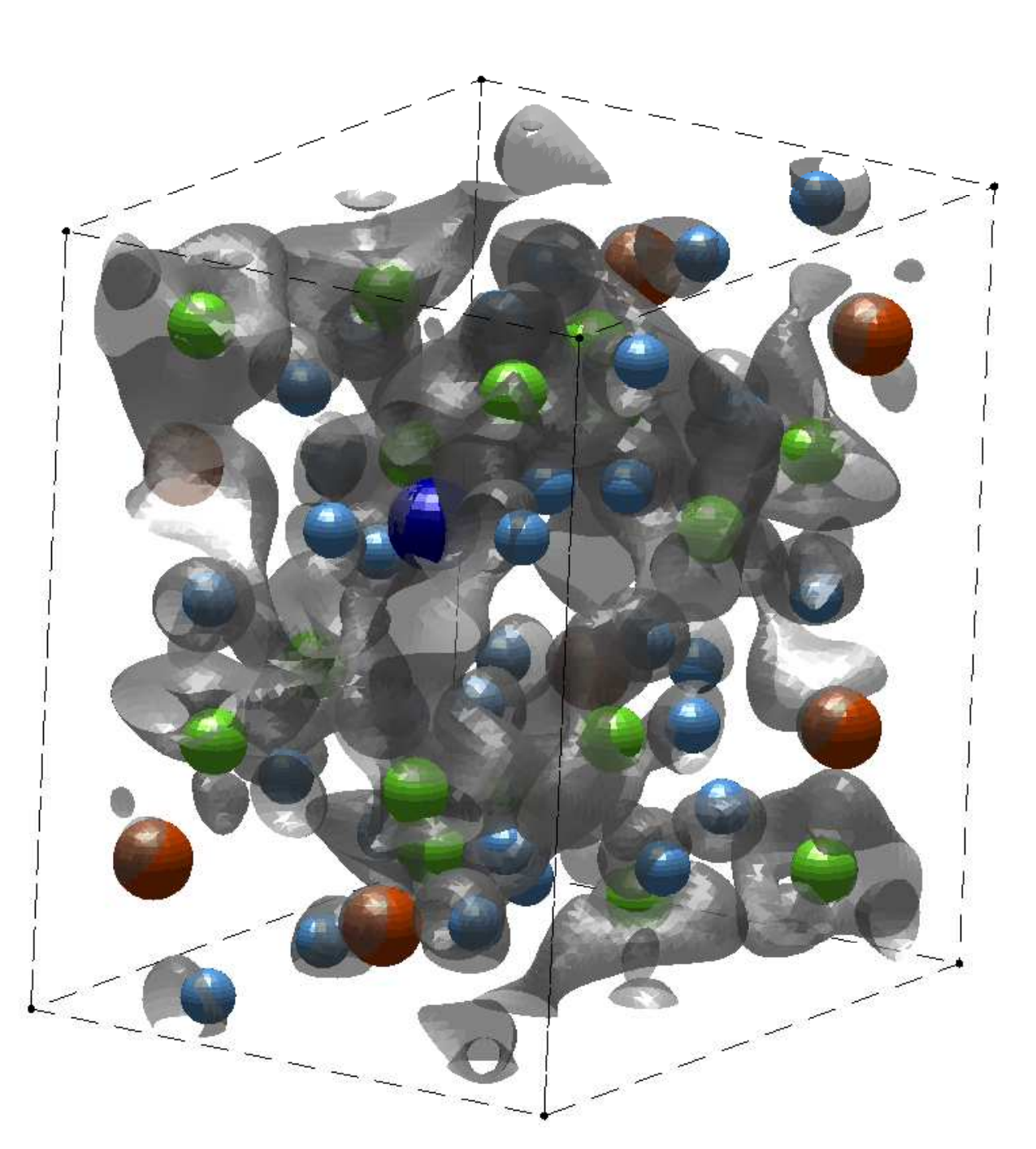}} %
\\
\subfloat[]{\label{fig:1c}\includegraphics[scale=0.43]{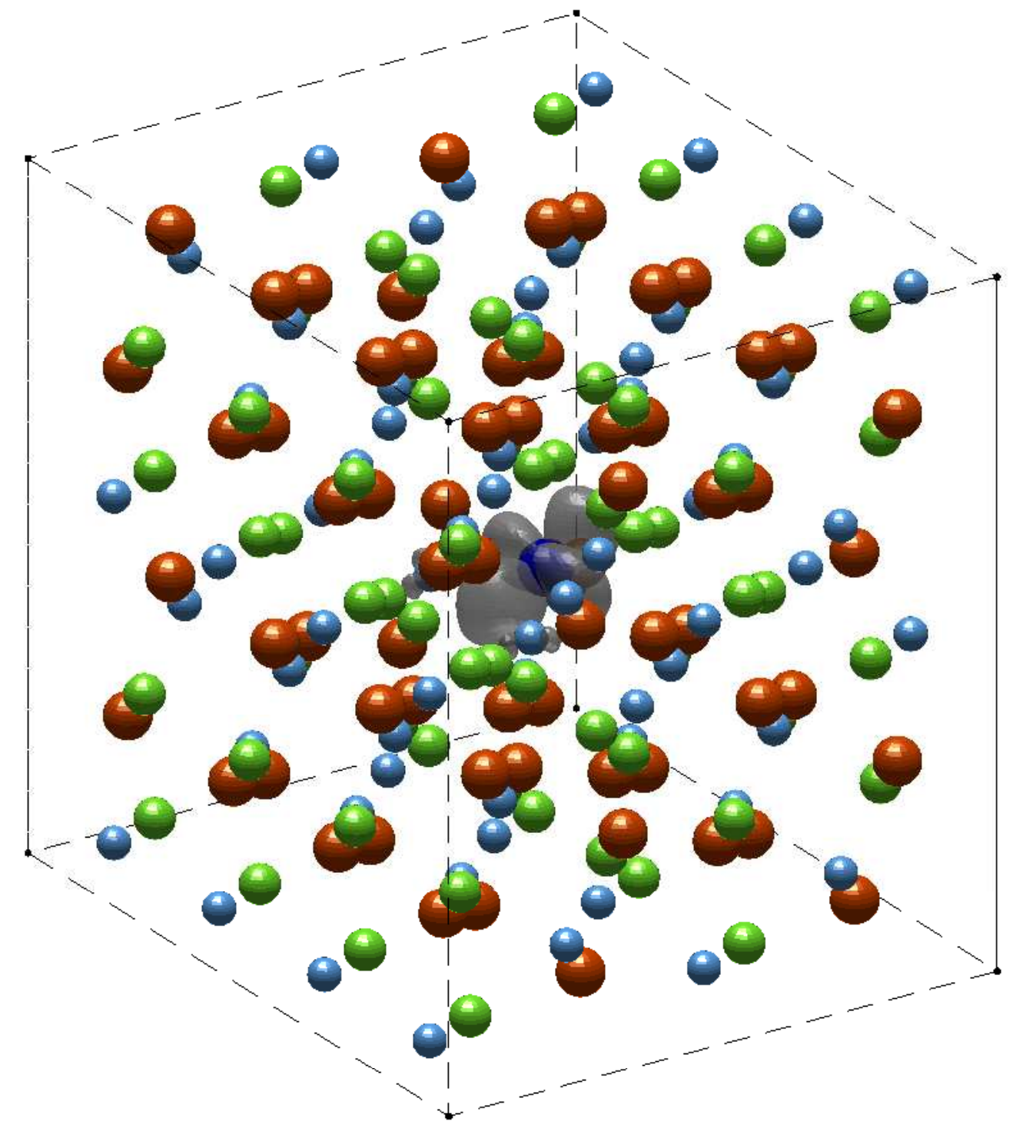}} \qquad
\subfloat[]{\label{fig:2b}\includegraphics[scale=0.39]{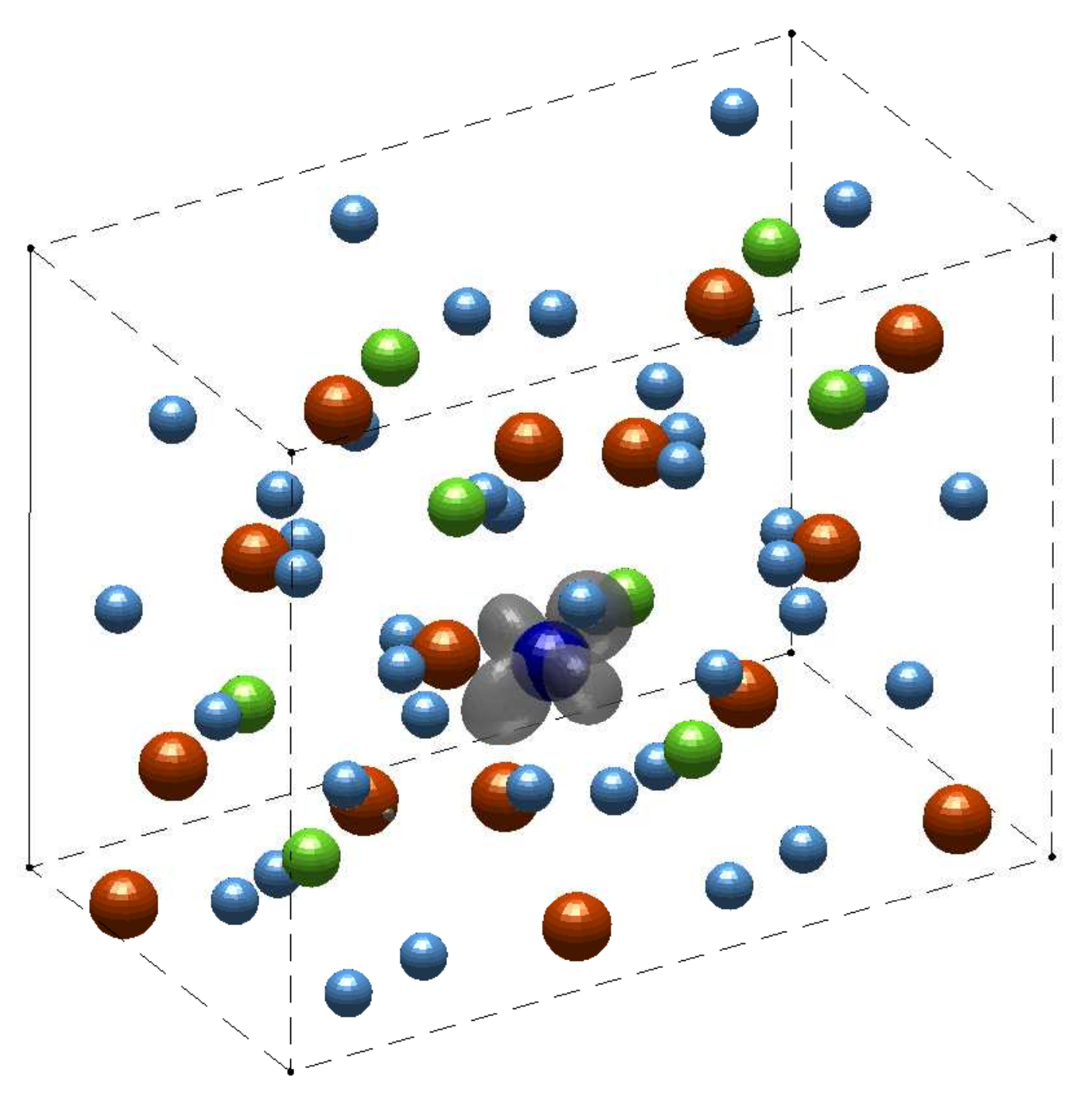}} \qquad   
\subfloat[]{\label{fig:2e}\includegraphics[scale=0.43]{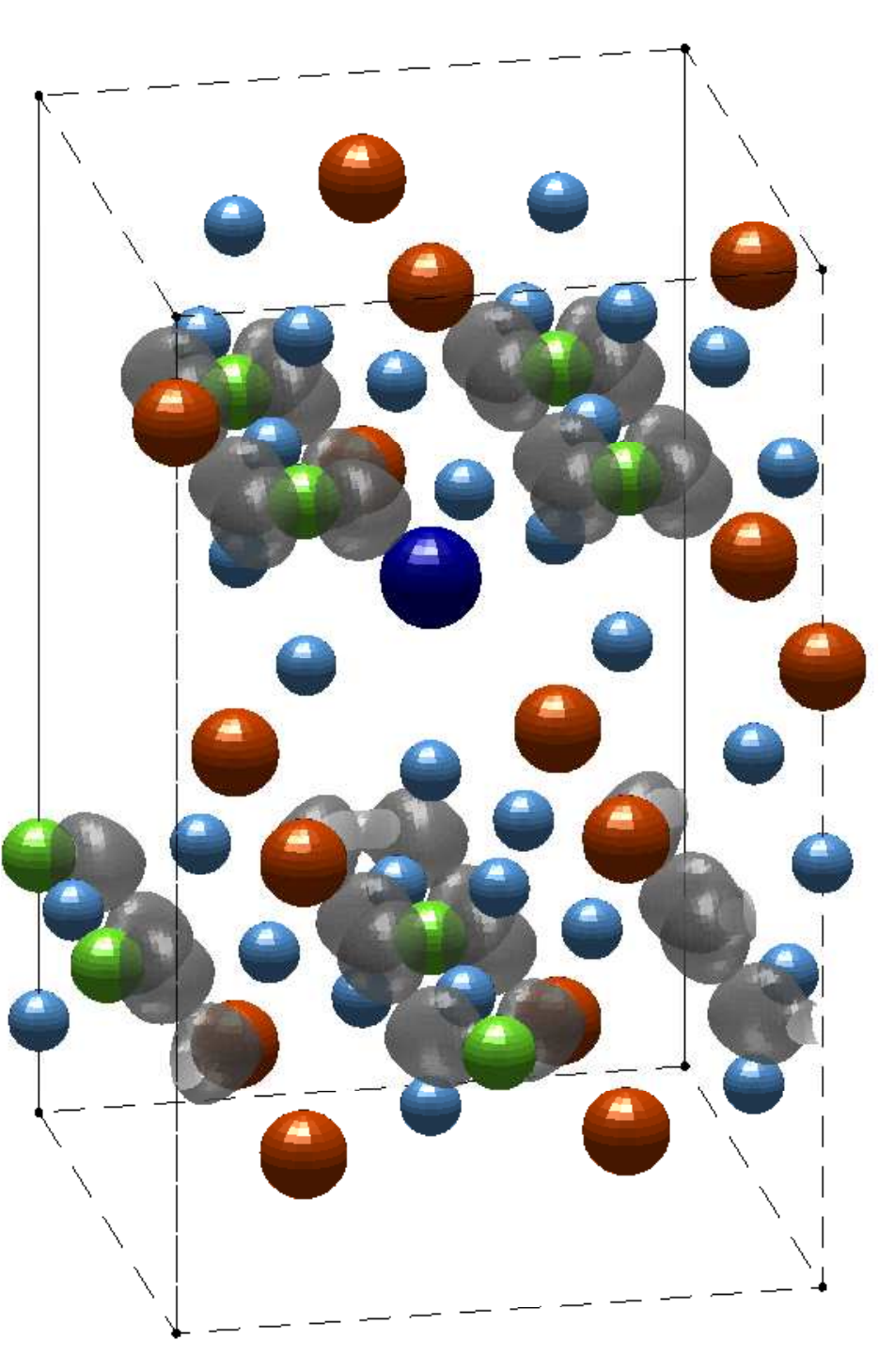}}    
\\
\includegraphics[scale=0.4]{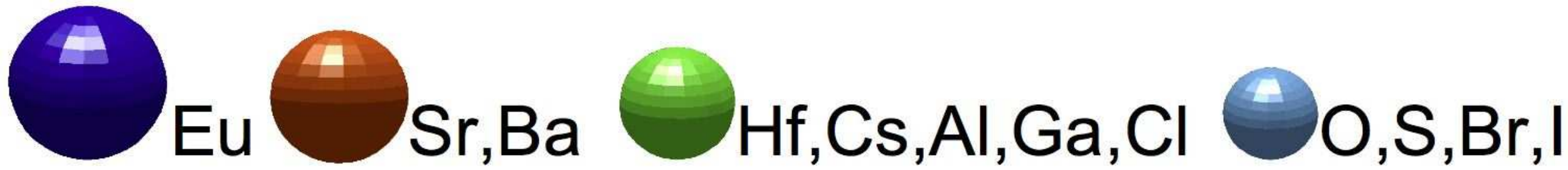}\\
\captionsetup{justification=justified}
\caption{\label{fig:fig2} Lowest excited state plots
for Eu doped scintillators and non-scintillators 
(a) SrI$_2$, (b) Ba$_2$AlS$_4$, 
(c) SrGa$_2$O$_4$, (d) BaClBr, (e) Ba$_2$CsI$_5$, (f) Ba$_2$HfS$_4$. Plots show the charge density 
isosurfaces of the excited state at 50\% threshold. The excited state is delocalized for
the non-luminescent compound SrGa$_2$O$_4$:Eu and has a predominantly Ga $s$ character while the excited state has $d$ character for other compounds. Ba$_2$HfS$_4$:Eu is a predicted non-scintillator wherein the first excited state has Hf 5$d$ character.}
\end{figure*}

Figure 2 shows the lowest  excited state for Eu-doped bright scintillators (SrI$_2$:Eu, BaAl$_2$S$_4$:Eu, Ba$_2$CsI$_5$:Eu, BaBrCl:Eu) and non-scintillators (SrGa$_2$O$_4$:Eu,  Ba$_2$HfS$_4$:Eu) to show the corresponding localized and delocalized nature of the excited states. Charge density iso-surfaces are plotted at 50\% threshold. The plots for the four bright scintillators show very atomic like Eu 5$d$ states while for the non-scintillators they represent states at the bottom of the host conduction band with Ga $4s$ character in the case of SrGa$_2$O$_4$:Eu and Hf 5$d$ character in the case of Ba$_2$HfS$_4$:Eu. 

The last four materials listed in Table II are our best new candidate materials, in terms of our theoretical criteria, for new bright scintillators from all the new systems we studied. 
For Ba$_2$CsBr$_5$:Eu the calculations were performed and it was theoretically predicted to be a good candidate for bright scintillation before it was synthesized in microcrystalline powder form\cite{borade2011}. For Ba$_2$CsI$_5$:Eu the calculations were performed and it was theoretically predicted to be a good candidate for bright scintillation before the successful synthesis and measurement of a microcrystalline powder sample with high luminosity. Ba$_2$CsI$_5$:Eu was then later grown in crystal form.\cite{bourret2009eu2} Ba$_2$CsI$_5$ and Ba$_2$CsBr$_5$ are isomorphous however, at the time of the calculations complete crystal structure parameters for Ba$_2$CsI$_5$ were not known. The calculations were done by using a relaxed crystal structure based on Ba$_2$CsBr$_5$, by replacing Br by I, which was later confirmed by X-ray data to be the correct structure.\cite{bourret2009eu2}

\section{\label{sec:conclusion}Conclusions}

A necessary condition for scintillation is that the Eu {4\textit f} and {5\textit d} levels should lie in the gap of the host material. The {4\textit f}-VBM and {5\textit d}-CBM energy gaps should be sufficient to ensure efficient hole trapping by Eu and avoid thermal quenching respectively. We did not find any examples of materials with Eu 4$f$ located below the host VBM. For all the systems where we found the Eu 5$d$ to be in the CB of the host material, to the best of our knowledge, there is no experimental evidence that any of them can be Eu activated. Therefore, similar to our studies on Ce-doped systems,\cite{Canning2011,chaudhry2011first} we did not find any examples of materials predicted to be non-scintillators but experimentally proven otherwise.

In terms of the different families of materials we found that for oxides and sulfides there were cases of systems where the Eu 5$d$ was in the CB while for the pure halides i.e. systems containing only Ca, Sr or Ba and one or two halides, the 5$d$ was always below the host CB in agreement with experimental results that all the known systems of this type show some level of activation with Eu. From a bandstructure point of view, compared to the sulfides and oxides, the halide family have the particular feature that even for the low bandgap systems, in particular the iodines, the 4$f$ and 5$d$ levels are still well placed in the gap of the host with the 5$d$ being far enough from the CBM to prevent thermal quenching and the 4$f$ being not too far from the VBM to facilitate hole trapping. In the case of oxides, for systems with a similar bandgap to the very bright iodine halides, the Eu 5$d$ level can be in the CB (see for example SrZrO$_3$ in Table~\ref{tab:table2}).  The major factor in the variation in scintillation luminosity for the halide systems is probably more related to competing non-radiative trapping processes on the host which are beyond the scope of this work.

In summary we have presented results for bandgaps, 4$f$ and 5$d$ levels and localization of the first excited state of Eu-doped scintillators and non-scintillators to relate theoretically calculable criteria to bright scintillation.  This approach was based on a method previously developed for Ce-doped systems\cite{Canning2011} and extended to Eu-doped systems. This approach has also allowed us to make qualitative predictions of candidates for new bright scintillators some of which have been successfully validated experimentally since our calculations were completed. 

\begin{acknowledgments}

We would like to thank Stephen Derenzo, Marvin J. Weber, Edith Bourret-Courchesne and Gregory Bizarri for many useful discussions 
and in particular Stephen Derenzo for help with candidate selection for the materials studied in our theoretical calculations.
We would also like to thank Pieter Dorenbos for providing the 4$f$-5$d$ absorption data listed in Table I and used in our calculations of 5$d$ levels.
The Materials Project was used as a resource in the research presented in this paper. \cite{Jain2013}
Work at the Lawrence Berkeley National Laboratory was supported by the 
U.S. Department of Homeland Security and carried out under U.S. Department of Energy Contract No.~DE-AC02-05CH11231.  

\end{acknowledgments}

%

\end{document}